\documentclass[12pt]{article}
\usepackage{graphicx}
\usepackage{type1cm, amsmath, amssymb}
\usepackage{cite}

\numberwithin{equation}{section}

\newcommand{\Cb}{\mathbb{C}}
\newcommand{\Rb}{\mathbb{R}}
\newcommand{\Ncal}{{\cal N}}
\newcommand{\del}{\partial}
\newcommand{\nn}{\nonumber}
\newcommand{\deldel}[2]{\frac{\del #1}{\del #2}}
\newcommand{\ox}{\otimes}
\newcommand{\id}{\mathbf{1}}
\newcommand{\Fcal}{{\cal F}}
\newcommand{\Ga}[1]{\Gamma_{\langle #1 \rangle}}
\newcommand{\Gi}{G_{\text{ind}}}
\newcommand{\vol}{{\rm vol}\,}
\newcommand{\im}{{\rm Im}\,}
\newcommand{\re}{{\rm Re}\,}
\newcommand{\FR}{F^{(5)}}
\begin{document}
\thispagestyle{empty}
\begin{flushright}
 \parbox{3.5cm}{UT-03-12 \\
  {\tt hep-th/0305007}}
\end{flushright}

\vspace*{2cm}
\begin{center}
 {\LARGE
 AdS Branes Corresponding to Superconformal Defects
 }
\end{center}

\vspace*{3cm} 
\begin{center}
 \noindent
 {\large Satoshi Yamaguchi}

 \vspace{5mm}
 \noindent
 \hspace{0.7cm} \parbox{130mm}{\it
 Department of Physics, Faculty of Science, University of Tokyo,\\
 Tokyo 113-0033, Japan.\\
 E-mail: {\tt yamaguch@hep-th.phys.s.u-tokyo.ac.jp}
}
\end{center}

\vspace{3cm}
\hfill{\bf Abstract\ \ }\hfill\ \\
We investigate an $AdS_4 \times L_2$ D5-brane in $AdS_5\times X_5$ space-time,
in the context of AdS/dCFT correspondence. 
Here, $X_5$ is a Sasaki-Einstein manifold and $L_2$ is a submanifold of $X_5$.
This brane has the same supersymmetry as the 3 
dimensional $\Ncal=1$ superconformal symmetry if $L_2$ is a {\em special 
Legendrian submanifold} in $X_5$. In this case, this brane is supposed to 
correspond to a superconformal wall defect in 
4-dimensional $\Ncal=4$ super Yang-Mills theory.
We construct these new string backgrounds
and show they have the correct supersymmetry, also in 
the case with non-trivial gauge flux on $L_2$. The simplest new example is 
$AdS_4\times T^2$ brane in $AdS_5\times S^5$.
We construct the brane solution expressing the RG flow between two
different defects.
 We also perform similar analysis for an $AdS_3\times L_3$ M5-brane
in $AdS_4 \times X_7$, for a weak $G_2$ manifold $X_7$ and its
submanifold $L_3$. This system has the same supersymmetry as 2-dimensional
$\Ncal=(1,0)$ global superconformal symmetry, if $L_3$ is an
{\em associative submanifold}.

\newpage
\section{Introduction}

Defect field theories appear in various fields in physics, and
an interesting problem.
Defect quantum field theories are useful in impurity
problem in condensed matter physics. 
Boundary conformal field theories are special class of defect field theories,
 and provide the celebrated worldsheet description of D-branes 
\cite{Polchinski:1995mt}.
In the string theory space-time, defect field theories
appear as the world-volume low energy theories in the intersecting brane systems
\cite{Sethi:1998zz,Ganor:1998jx,Kapustin:1998pb}.

AdS brane/defect CFT (AdS/dCFT) correspondence 
proposed in \cite{Karch:2000ct,Karch:2000gx} 
is the approach from the AdS/CFT correspondence to these defect field theories.
Various aspects of the AdS/dCFT 
correspondence have been investigated in  
\cite{Bachas:2001vj,DeWolfe:2001pq,Erdmenger:2002ex,Skenderis:2002vf,%
Karch:2002sh,Mateos:2002bu,Yamaguchi:2002pa,Constable:2002xt,%
Constable:2002vt,Aharony:2003qf,Suryanarayana:2003}. 
The most typical example of the AdS/dCFT is the type IIB one. The string theory
side of the correspondence is the IIB supergravity on $AdS_5\times S^5$
with $AdS_4 \times S^2$ brane whose effective theory is the Dirac-Born-Infeld 
action. The field theory side is $\Ncal=4$ super 
Yang-Mills theory with the wall defect on which the fundamental hyper-multiplet
lives \cite{Sethi:1998zz,Ganor:1998jx,Kapustin:1998pb,DeWolfe:2001pq}.

Until now, the $AdS_m \times S^{n}$ branes
and their corresponding defects have been mainly investigated,
but $AdS_{m} \times L_{n}$ branes with non-spherical $L_{n}$
have been less investigated. An $AdS_m \times L_n$ brane seems to correspond
to a nontrivial conformal fixed point of the defect field theory. 
The RG flows between these fixed points and their brane pictures
are good phenomena to see the correspondence.

We study, in this paper, rather general $AdS_{m} \times L_{n}$ type branes.
In IIB string theory, we consider an $AdS_4\times L_2$ brane
in $AdS_5\times X_5$ space-time,  where $X_5$ is a Sasaki-Einstein manifold.
We show that if $L_2$ is a special Legendrian submanifold in $X_5$, the 
background preserves the same supersymmetry as 3-dimensional $\Ncal=1$ 
superconformal symmetry, as expected.
In this analysis, we treat a bent D-brane in AdS
part with appropriate gauge flux in $L_2$.
These are treated in \cite{Karch:2000gx,Skenderis:2002vf}
in the $L_2=S^2$ case. In the case with non-zero flux,
the ambient theories of left and right side of the defect are distinct. 

The most simple non-trivial example of the special Legendrian submanifold
is appropriately embedded $T^2$.
We investigate $AdS_4\times T^2$ brane and its corresponding
defect CFT. Especially, this CFT can flow to the corresponding CFT
of $AdS_4\times S^2$ brane. We construct the solution of the flow in the
brane picture.

We also consider $AdS_3\times L_3$ M5-brane in $AdS_4\times X_7$ space-time,
where $X_7$ is a weak $G_2$ manifold. We show that if $L_3$ is an associative
submanifold, this background has the same supersymmetry as 2-dimensional global
superconformal symmetry, as expected. In this analysis, the D-brane can bend
in $AdS_4$ and admit appropriate 3-form flux at the same time.
In this system, the ambient theories of left and right
of the defect are distinct as suggested in 
\cite{Karch:2000gx} in the case with $L_3=S^3$ in $AdS_4 \times S^7$.

The construction of this paper is as follows: Section \ref{IIB} discuss
the supersymmetry of IIB $AdS_4\times L_2$ brane
in $AdS_5\times X_5$. In section \ref{T2}, we treat the $AdS_4\times T^2$ brane,
and its flow to $AdS_4\times S^2$ brane.
In section \ref{M}, we consider the supersymmetry of $AdS_3\times L_3$
M5-brane in $AdS_4\times X_7$ background of M-theory.
Section \ref{Conclusion} is devoted to conclusions and discussions.
We write some definitions and the proof of some
formulas in the appendix.

\section{IIB $AdS_4\times L_2$ D5-brane in $AdS_5\times X_5$}
\label{IIB}
In this section, we study the remaining supersymmetry in the presence of
the $AdS_4\times L_2$ D5-brane in $AdS_5\times X_5$ space-time of IIB theory.
In this paper, we consider the general Sasaki-Einstein manifold $X_5$,
and the general special Legendrian submanifold $L_2$ in $X_5$. We also
includes the gauge flux on the brane in $L_2$ that makes the AdS brane bend.
We show that there are the same amount of the supersymmetry as 3-dimensional
superconformal symmetry as expected, by using the probe approximation.
The analysis in this section is generalisation of
the analysis of $AdS_4\times S^2$ brane in \cite{Skenderis:2002vf} to 
general special Legendrian submanifold $L_2$. 

In order to perform this analysis, we first review the construction of
the Killing spinors in $AdS_5\times X_5$ background.
This type of backgrounds have been investigated in
\cite{Romans:1985an,Klebanov:1998hh,Gubser:1998vd,Acharya:1998db,Morrison:1998cs}.
Next, we use the kappa symmetry projection to determine
the surviving supersymmetry in the presence of $AdS_4\times L_2$ D5-brane.

\subsection{Supersymmetry of the closed string background}

Let us first describe the $AdS_5\times X_5$ solution in 10 dimensional
IIB supergravity and
fix the convention. We only turn on the metric and the 5-form
field strength $\FR$ in IIB supergravity.
The Einstein equation can be written as
 \begin{align}
 R_{MN}=\frac14 \FR_{MPQRS}\FR_{N}{}^{PQRS},\qquad M,N,\dots=0,\dots, 9.
\end{align}
The Bianchi identity for $\FR$ and self-duality are also required. 
The metric for the $AdS_5\times X_5$ is described as
\begin{align}
& ds^{2}=dr^2+e^{-2r}dx^{\mu}dx^{\nu}\eta_{\mu\nu}+g_{mn}dy^{m}dy^{n},\nn\\
& \mu,\nu=0,1,2,3,\qquad m,n=5,6,7,8,9,\qquad
 \eta_{MN}={\rm diag}(-1,1,1,\dots),\label{IIB-metric}
\end{align}
where $g_{mn}$ is the Sasaki-Einstein metric of $X_5$ normalised as
\begin{align}
 R_{mn}=4g_{mn}.\label{Unity}
\end{align}
We can always rescale $g_{mn}$ so that Eq.(\ref{Unity}) is satisfied, if
$X_5$ is an Einstein manifold with a positive cosmological constant.
We set the vielbein $e^{a'},\ a'=0,1,2,3,4$ of $AdS_5$ space-time as
\begin{align}
 e^{p}:=e^{-r}dx^{p},\ (p=0,1,2,3),\qquad e^{4}:=dr,
\end{align}
and we denote by $e^{a},\ (a=5,\dots,9)$ 
the vielbein for the metric $g_{mn}$ of
$X_5$ space. In this notation, the solution of 5-form can be written as
\begin{align}
 \FR=4(e^{0}e^{1}e^{2}e^{3}e^{4} + e^{5}e^{6}e^{7}e^{8}e^{9}). \label{IIB-5-form}
\end{align}
Actually, the solution (\ref{IIB-metric})(\ref{IIB-5-form}) can have a parameter;
the radius of $AdS_5$ (or $X_5$). In this paper, however, we set the radius
to be $1$ because it is irrelevant in the analyses below.

Next, we turn to the Killing spinors. The supersymmetry condition of
the gravitino in the background of the metric and the 5-form is expressed as
\begin{align}
  \del_{A}\epsilon+\frac14 \omega_{A}{}^{BC}\Gamma_{BC}\epsilon
  +\frac{i}{2^4\times 5!}\FR_{BCDEF}\Gamma^{BCDEF}\Gamma_{A}\epsilon=0,\qquad
 \del_{A}:=(e^{-1})^{M}_{A}\deldel{}{x^{M}},\nn\\
(A,B,\dots=0,1,\dots,9),\label{Killing-spinor1}
\end{align}
where $\epsilon=\epsilon_{L}+i\epsilon_{R}$ is the parameter of the 
supersymmetry, and $\epsilon_{L,R}$ 's are Majorana-Weyl spinors with positive
chirality. The dilatino condition is trivially satisfied in this case.
If we insert the form (\ref{IIB-5-form}) to eq.(\ref{Killing-spinor1}),
we obtain the form
\begin{align}
  \del_{A}\epsilon+\frac14 \omega_{A}{}^{BC}\Gamma_{BC}\epsilon
  +\frac{i}{2}\Gamma^{01234}\Gamma_{A}\epsilon=0. \label{Killing-spinor2}
\end{align}
In this equation, the torsionless spin connection of $AdS_5$ becomes 
$\omega^{p4}=-\omega^{4p}=-e^{p},\ (p=0,1,2,3)$ and other components are $0$.
We also denote the spin connection of $X_5$ by $\omega^{ab},\ (a,b=5,\dots,9)$
which satisfies $de^{a}+\omega^{ab}e^{b}=0$.

To solve the equation (\ref{Killing-spinor2}), we decompose $\epsilon$ in the 
same way as \cite{Claus:1998yw}
\begin{align}
 \epsilon=\epsilon_{+}+\epsilon_{-},\qquad
 i\Gamma^{0123}\epsilon_{\pm}=\pm \epsilon_{\pm}.
\end{align}
Then, eq.(\ref{Killing-spinor2}) becomes
\begin{align}
  &\deldel{}{r}\epsilon_{\pm}\pm \frac12 \epsilon_{\pm}=0,\nn\\
 &\deldel{}{x^p}\epsilon_{+}+e^{-r} \Gamma_{4p}\epsilon_{-}=0,\qquad
 \deldel{}{x^p}\epsilon_{-}=0,\qquad p=0,1,2,3, \nn\\
 &\del_{a}\epsilon_{\pm}+\frac14 \omega_{a}{}^{bc}\Gamma_{bc}\epsilon_{\pm}
  \pm\frac 12 \Gamma_{4a}\epsilon_{\pm}=0,\qquad a,b,c=5,\dots,9.
 \label{Killing-spinor3}
\end{align}
It is convenient to take the 10 dimensional gamma matrices $\Gamma_A,\ \{\Gamma_A,\Gamma_B\}=2\eta_{AB}$ as the
following form.
\begin{align}
  \Gamma^{a'}=\gamma^{a'}\ox \id \ox \sigma_{1},
 \qquad \Gamma^{a}=\id \ox \gamma^{a} \ox \sigma_{2},
 \qquad \Gamma^{11}=\id \ox \id \ox \sigma_{3},
\label{IIB-gamma}
\end{align}
where $\gamma^{a'},\ a'=0,1,2,3,4$ are $4\times 4$ gamma matrices for SO(4,1),
and $\gamma^{a},\ a=5,\dots,9$ are $4\times 4$ gamma matrices for SO(5). These 
gamma matrices satisfy the relations
\begin{align}
  \{\gamma^{a'},\gamma^{b'}\}=\eta^{a'b'},\qquad
 \{\gamma^{a},\gamma^{b}\}=\delta^{ab},\qquad
 \gamma^{4}=i\gamma^{0123},\qquad \gamma^{56789}=\id.
\end{align}
In eqs.(\ref{IIB-gamma}), we also use the Pauli matrices $\sigma_{j},\ j=1,2,3$.
By using these notations, the equation (\ref{Killing-spinor3}) can be solved as
\begin{align}
  &\epsilon=\lambda \ox \chi_{-} \ox \binom{1}{0},\\
 &\lambda=e^{\frac12 r}\zeta_{-}+e^{-\frac12 r}\left(x^{p} \gamma_{p4} \zeta_{-}
     +\zeta_{+}\right),\qquad p=0,1,2,3,\\
 &\zeta_{\pm}: (\text{4 component constant spinor}),\qquad
   i\gamma^{0123}\zeta_{\pm}=\pm \zeta_{\pm},\\
 &\del_{a}\chi_{-}+\frac14 \omega_{a}{}^{bc}\gamma_{bc}\chi_{-}
  +\frac{i}{2}\gamma_{a}\chi_{-}=0. \label{Real-Killing-spinor}
\end{align}
Eq.(\ref{Real-Killing-spinor}) implies that $\chi_{-}$ is a real Killing spinor
in $X_5$.
It is known that if and only if $X_5$ is a Sasaki-Einstein manifold, there is
a real Killing spinor \cite{Bar:1993}. 

In summary, we write down the Killing spinors of $AdS_5\times X_5$ in this 
subsection. These Killing spinors exist only when $X_5$ is a Sasaki-Einstein
manifold. We consider $AdS_4\times L_2$ D5-branes in this background
and their supersymmetry in the next subsection.

\subsection{Supersymmetry of D-brane background}

Let us introduce an $AdS_4\times L_2$ D5-brane into the background
considered in the previous subsection. We show that this system has
the same amount of the supersymmetry as the 3-dimensional $\Ncal=1$ 
superconformal symmetry, as expected. 

As in \cite{Cederwall:1997pv,Aganagic:1997pe,Cederwall:1997ri,%
Bergshoeff:1997tu,Aganagic:1997nn,Bergshoeff:1997kr,Skenderis:2002vf},
the surviving Killing spinors when we put the
D-brane should satisfy
\begin{align}
 \Gamma \epsilon = \epsilon,
\end{align}
where $\Gamma$ is the matrix used for the kappa symmetry 
projection. The matrix $\Gamma$ for a IIB D$p$-brane can be written as
\begin{align}
 &d^{p+1}\xi \Gamma = -e^{-\Phi}(-\det(\Gi + \Fcal))^{-1/2} e^{\Fcal}
     {\cal X}|_{(p+1)\text{-form}}, \label{Gamma1}\\
 &{\cal X}:=\sum_{n} \frac{1}{(2n)!}d\xi^{i_{2n}}\dots d\xi^{i_{1}}
 \Ga{i_1i_2\dots i_{2n}} K^{n} (-i),\\
 &\Ga{i_1i_2\dots i_{s}}:=\deldel{X^{M_1}}{\xi^{i_1}}\dots \deldel{X^{M_{s}}}{\xi^{i_{s}}}
 e_{M_1}^{A_1}\dots e_{M_{s}}^{A_{s}}\Gamma_{A_{1}\dots A_{s}},\label{Gamma2}
\end{align}
where $\xi^{i}$ are the world-volume coordinates, $\Phi$ is the dilaton,
$\Gi$ is the
induced metric, $\Fcal=F-B$ is the linear combination of the NSNS B-field and
the world-volume gauge field, and $K$ is the charge conjugation
$K\epsilon=\epsilon^{c}$.

The brane configuration treated here is described as follows. First, we set
the world-volume coordinate as $\xi^{i},\ i=0,1,2,4,5,6$, and take the
static gauge (use the space-time coordinate in eq.(\ref{IIB-metric}))
\begin{align}
 x^{j}=\xi^{j},\quad (j=0,1,2),\qquad r=\xi^{4}. \label{IIB-static-gauge}
\end{align}
We use $x^{0,1,2},r$ as both space-time and world-volume coordinates.
Secondly, we consider the situation where the $AdS$ part of the D-brane
bend as $x^{3}=M e^{r}$, where $M$ is a constant. Thirdly, the
immersion $(\xi^5,\xi^6)\to X_5$ is a special Legendrian immersion, and
its image is $L_2$.
Then the induced metric becomes
\begin{align}
& (\Gi)_{ij}d\xi^{i}d \xi^{j}
  =(1+ M^2) dr^2 + e^{-2r} dx^{\alpha}dx^{\beta}\eta_{\alpha\beta}
 + G_{pq}d\xi^{p} d\xi^{q},\qquad(\alpha,\beta=0,1,2,\quad p,q=5,6),\nn\\
&G_{pq}:=g_{mn}\deldel{y^m}{\xi^{p}}\deldel{y^n}{\xi^{q}},
\end{align}
where $G_{pq}$ is the induced metric of $L_2$. Note that
through this section, we use $G$ as the induced metric of $L_2$. The total
induced metric are denoted by $\Gi$.
Finally, we introduce the world-volume gauge field excitation
$\Fcal=f \sqrt{\det G}d\xi^{5}d\xi^{6}$ , where $f$ is a constant.
Then the DBI determinant becomes
\begin{align}
  -\det(\Gi + \Fcal)=(1+M^2)e^{-6r}\det (G+\Fcal)
 =(1+M^2)(1+f^2) e^{-6r} \det G.
\end{align}
In the above D-brane configuration, the matrix $\Gamma$ in
eq.(\ref{Gamma1})-(\ref{Gamma2}) becomes
\begin{align}
  \Gamma=\left[
 (1+M^2)(1+f^2) \det G\right]^{-1/2}
i\Gamma_{012}(\Gamma_{4}+M\Gamma_{3})\left[
 \Ga{56} K
 + f\sqrt{\det G}
  \right].
\end{align}

Now, let us consider the equation $\Gamma \epsilon=\epsilon$ for the Killing 
spinors $\epsilon$. We will show that half of the $\epsilon$ satisfy this
equation when we set appropriate relation between $f$ and $M$. The key
formula for this analysis is
\begin{align}
 \deldel{y^m}{\xi^{5}}\deldel{y^n}{\xi^{6}} e^{a}_{m}e^{b}_{n}\gamma_{ab}\chi_{-}^{c}
   =\sqrt{\det G}\chi_{-},\label{IIB-key-formula}
\end{align}
for a special Legendrian submanifold $L_2$ and a real Killing spinor
$\chi_{-}$ of $X_5$ with certain phase. We will show
eq.(\ref{IIB-key-formula}) in appendix \ref{app-IIB-key-formula}.
If we use eq.(\ref{IIB-key-formula}), what we should show is
\footnote{We use here the convention
$(\gamma_{a'}\zeta)^{c}=-\gamma_{a'}\zeta^{c}$ for SO(3,1) spinor $\zeta$.}
\begin{align}
 &\lambda=\Gamma'\lambda := \frac{1}{\sqrt{
 (1+M^2)(1+f^2)}}(\gamma_3-\gamma_4M)(\lambda^{c}+ f \lambda),\nn\\
 &\lambda:=e^{-\frac12 r}(-x^{\alpha}\gamma_{\alpha}\zeta_{-}+ \zeta_{+})
   +e^{+\frac12 r}(1-M\gamma_{3})\zeta_{-},\nn\\
 &\lambda^{c}:=e^{-\frac12 r}(x^{\alpha}\gamma_{\alpha}\zeta_{-}^{c}
    + \zeta_{+}^{c})
   +e^{+\frac12 r}(1+M\gamma_{3})\zeta_{-}^{c}.
\end{align}

$\Gamma'\lambda$ can be expanded as
\begin{align}
 \Gamma'\lambda=\frac{1}{\sqrt{(1+M^2)(1+f^2)}}
 \Bigg[&
    e^{-\frac12 r}\Big\{
       -x^{\alpha}\gamma_{\alpha}(\gamma_3\zeta_{-}^{c} - M \zeta_{-}^{c}
  -f \gamma_{3} \zeta_{-} - f M \zeta_{-})\nn\\
  &\qquad\qquad
  +(\gamma_3\zeta_{+}^{c}+M\zeta_{+}^{c}+f\gamma_3\zeta_{+}-fM\zeta_{+})
    \Big\}\nn\\
  &+e^{\frac12 r}(1+M^{2})(\gamma_3 \zeta_{-}^{c}+f \gamma_{3}\zeta_{-})
 \Bigg].
\end{align}
From this equation, we obtain the following results. First, if we want a 
supersymmetry, the bending $M$ and the gauge flux $f$ must satisfy the relation
$f=-M$. Next, if this relation is satisfied, the surviving Killing spinors
are the ones with
 $ \gamma_{3}\zeta_{-}^{c}=\zeta_{-},\quad \gamma_{3}\zeta_{+}^{c}=\zeta_{+}$.
The number of remaining supercharges is $4$ in general. This is the same 
number as the 3-dimensional $\Ncal=1$ superconformal symmetry.

Let us check this brane configuration actually satisfies the field equation.
The bosonic part of the D5-brane action on this background is
\begin{align}
 I=-\int d^6\xi \sqrt{-\det(\Gi+\Fcal)}-\int \Fcal \wedge C^{(4)},
\end{align}
where $C^{(4)}$ is the RR 4-form potential $d C^{(4)}=\FR$.
If we take the static gauge (\ref{IIB-static-gauge}), the remaining fields are
$x^{3},\ y^{m},\ {\cal A}$; ${\cal A}$ is the world-volume gauge field
$\Fcal=d{\cal A}$. The $y^m$ and ${\cal A}$ are easily checked to satisfy
the field equation. Note that we use here the fact that a special Legendrian 
submanifold is a minimal submanifold
\footnote{By the term ``minimal submanifold'',
we express that the variation of the volume vanishes under
small fluctuation. It is not necessarily a submanifold with minimal volume.}. 
The equation of motion for $x^3$,
after we insert the solution $\Fcal$ and $y^m$ and anzats $x^{3}=x^{3}(r)$,
becomes
\begin{align}
 \deldel{}{r}\left[\sqrt{(1+M(r)^2)(1+f^2)\det G} e^{-5r}(1+M(r)^2)^{-1}\deldel{x^3}{r}\right]
-4f\sqrt{\det G}e^{-4r}=0,\\
 M(r):=e^{-r}\deldel{x^3}{r}.
\end{align}
We can easily check that $x^{3}(r)=-fe^{r}$ is a solution
of this equation of motion.

\section{$AdS_4\times T^2$ brane and its flow to $AdS_4\times S^2$}
\label{T2}
In this section, we explain a nontrivial
example of special Legendrian submanifolds in $S^5$:
appropriately embedded $T^2$.
Many examples of special Legendrian submanifolds in $S^5$ (which has one to one
correspondence to special Lagrangian cones in $\Cb^3$) are shown in 
\cite{Hearvey:1982,Joyce:2001xt}. Among them, we consider here the simplest
non-trivial one. In this section, we limit ourselves to the case with
$M=f=0$ for simplicity.

The precise immersion of this $T^2 \to S^5$ are described as follows.
We parametrise $S^{5}$ by $(z_1,z_2,z_3)\in \Cb^3,\quad 
|z_1|^2+|z_2|^2+|z_3|^2=1$. Then, the special Legendrian $T^2$ can be expressed
as
\begin{align}
 z_j=\frac{1}{\sqrt 3}e^{i\theta_j},\quad j=1,2,3,\qquad
 \theta_1+\theta_2+\theta_3 \equiv 0 \mod{2\pi}.
\end{align}

The corresponding defect CFT is expected as follows. First consider D3-D5 system
described in table \ref{table1}.
\begin{table}
\begin{center}
  \begin{tabular}{|c||c|c|c|c|c|} \hline
  &0 &1 &2 &3 &4--9 \\ \hline
 D3 &$\bigcirc$ &$\bigcirc$ &$\bigcirc$ &$\bigcirc$ & $\times$ \\ \hline
 D5 &$\bigcirc$ &$\bigcirc$ &$\bigcirc$ &$\times$ & $\tilde L$\\ \hline
 \end{tabular}
\end{center} 
\caption{The D3-D5 configuration considered here. ``$\bigcirc$''
 means ``extended'' and ``$\times$''
 means ``not extended''. The D5-brane is wrapped on the special 
Lagrangian submanifold $\tilde L$ in 4--9 directions.}
 \label{table1}
\end{table}
The number of D3-branes is $N$, and that of D5-brane is one.
In 4--9 directions, the D5-brane is wrapped on a special Lagrangian
submanifold $\tilde L$; This $\tilde L$ is described
as the cone over the special Legendrian $T^2$.
On the other hand, all $N$ D3-branes sit at the tip (singularity) of $\tilde L$
in 4--9 direction. Consider the near horizon region of the
supergravity solution of D3-branes in this configuration. 
In this region, the space-time becomes
$AdS_5\times S^5$ and the D5-brane (treated as a probe)
becomes $AdS_4\times T^2$ brane in this 
space-time. This is the background of
the string theory side of the correspondence.

Next, the theory of the field theory side will be
obtained as the low energy theory 
on the D3-branes. This theory includes 2 sectors: 3-3 string sector and 3-5 
string sector. Low energy theory of 3-3 string sector is 4-dimensional $\Ncal=4$
U($N$) super Yang-Mills theory as usual.
The 3-5 string sector and its coupling to 3-3
string sector characterise the defect conformal field theory. Because
the D3-branes sit at the singularity of the D5-brane, the theory of 3-5 string 
is singular. This makes it difficult to describe the low energy theory.
We postpone this analysis to the future work.

Now, we have two kinds of defects in the ambient theory $\Ncal=4$ super 
Yang-Mills theory; one (we call it ``$S^2$ defect'')
 corresponds to $AdS_4\times S^2$ brane, and the other (we call it ``$T^2$
 defect'') corresponds to $AdS_4\times T^2$ brane. Let us consider here
the relation between $S^2$ defect and $T^2$ defect. If there is 
some relation between these two defects, it might give a hint on
the description of $T^2$ defect since the theory of $S^2$ defect
is known \cite{Sethi:1998zz,Ganor:1998jx,Kapustin:1998pb,DeWolfe:2001pq}.

In order to see the renormalization group flow of $S^2$ defect and $T^2$ defect,
we consider here the g-function in the point of view of the $AdS$ 
brane\cite{Yamaguchi:2002pa}. In the present case, the g-function at
each fixed point is proportional to the area of $S^2$ or $T^2$.
The ratio of g-function of $T^2$ defect ($g_{T^2}$) and that of $S^2$ defect 
($g_{S^2}$) becomes
\begin{align}
 \frac{\ln g_{T^2}}{\ln g_{S^2}}=\frac{\pi}{\sqrt{3}} > 1.
\end{align}
Therefore, $g_{T^2} > g_{S^2}$ is satisfied and the RG flow from $S^2$ defect
to $T^2$ defect is forbidden. On the other hand, the RG flow from $T^2$ defect
to $S^2$ defect may exist. In the rest of this section, we show that
this $T^2 \to S^2$ flow can be described in the $AdS$ brane side.

The key of this flow is the existence of the
special Lagrangian manifold which is non-singular and asymptotically $T^2$ cone.
This special Lagrangian submanifold $\tilde L_{\mu}$
in $\Cb^3=\{(z_1,z_2,z_3)\}$ can be described as
\begin{align}
 z_2=\bar{z}_1 e^{-i\alpha},\qquad
 z_3=\sqrt{|z_1|^2+\mu^2}e^{i\alpha}-\mu,\qquad
 z_1\in \Cb,\qquad e^{i\alpha} \in S^1,
 \label{SL}
\end{align}
where $\mu$ is a real positive deformation parameter.
In the region $|z_1| \gg \mu$, $\tilde L_{\mu}$ asymptotically reach $\tilde L
= \tilde L_{\mu=0}$. On the other hand, around the origin, this submanifold
is smooth and can be approximated by a plane.

\begin{figure}
 \begin{center}
  \begin{tabular}{ccc}
     \includegraphics{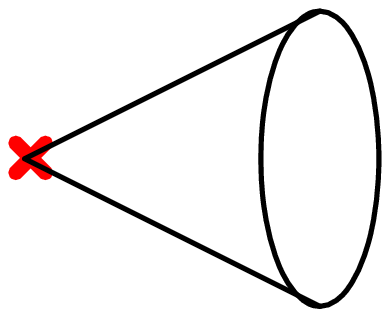}
     &\includegraphics{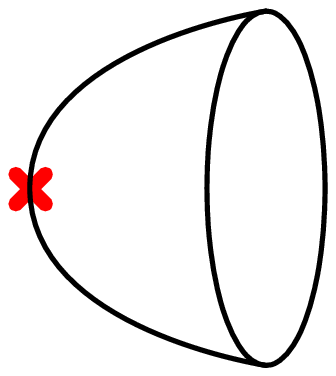}
     &\includegraphics{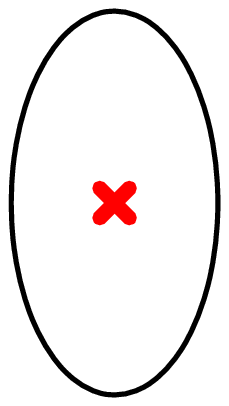}\\
   (a) & (b) & (c)
  \end{tabular} 
\end{center}
 \caption{The image of the flow from $T^2$-brane to $S^2$-brane.
 (a) UV limit: The D3-branes are sitting at the tip of the $T^2$ cone.
 In the near horizon limit, the D5-brane becomes $AdS_4\times T^{2}$ and
 corresponds to $T^2$ defect.
 (b) Intermediate scale: The defect theory has the RG flow.
 (c) IR limit: The D3-branes are sitting on a flat D5-brane.
 In the near horizon limit, the D5-brane becomes $AdS_4\times S^{2}$ and
 corresponds to $S^2$ defect.
 }
\label{fig1}
\end{figure}

Let us introduce large number of D3-branes at the origin of $\tilde L_{\mu}$ and
a D5-brane extended to the direction 0,1,2 and $\tilde L_{\mu}$.
Consider the near horizon geometry D3-branes. The bulk geometry is
$AdS_5\times S^5$ described by the metric
\begin{align}
 ds^{2}=t^{2}dx^{\mu}dx^{\nu}\eta_{\mu\nu} + t^{-2}(\sum_{j=1}^{3}|dz_j|^2),
 \label{AdSxS}
\end{align}
where $t=\sqrt{\sum_{j=1}^{3}|z_j|^2}$. If we define $t=e^{-r}$ and
introduce $S^5$ coordinates $y^{m}$, the metric is expressed by 
eq.(\ref{IIB-metric}). In the UV region (negatively large 
$r$ or large $t$), $\tilde L_{\mu}$ becomes
the cone over $T^2$, and the D5-brane
in eq.(\ref{AdSxS}) looks like $AdS_4\times T^{2}$. 
 In the IR region (positively large 
$r$ or small $t$), $\tilde L_{\mu}$ becomes a plane (cone over $S^2$),
and the D5-brane in eq.(\ref{AdSxS}) looks like $AdS_4\times S^{2}$. 
Consequently, this brane configuration expresses the flow from
$T^2$ defect to $S^2$ defect. The image of this flow is illustrated in
figure \ref{fig1}.

We can show that there is expected amount of supersymmetry in
the $AdS_5 \times S^5$ geometry of eq.(\ref{AdSxS})
with D5-brane described with eq.(\ref{SL}) and extending to 012 direction.

\section{$AdS_3\times L_3$ M5-brane in $AdS_4\times X_7$}
\label{M}
In this section, we perform the similar analysis as section \ref{IIB}
for the $AdS_3\times L_3$ M5-brane in $AdS_4\times X_7$ space-time
of M-theory.
The M-theory background $AdS_4\times X_7$ is a typical example of Freund-Rubin
compactifications \cite{Freund:1980xh}. If $X_7$ is a weak $G_2$ manifold,
this background has the supersymmetry. For a review of this type of 
compactifications, see \cite{Duff:1986hr}.
These backgrounds have been also investigated in
 \cite{Acharya:1998db,Morrison:1998cs} in the context of AdS/CFT correspondence.

In this paper, we consider M5-brane wrapped on $AdS_3\times L_3$ in above 
background. This brane background corresponds to a defect CFT: 3-dimensional
CFT with scale invariant wall defect. We show that if $L_3$ is an associative
submanifold in $X_7$, the brane background has the same supersymmetry as
2-dimensional $\Ncal=(1,0)$ global superconformal symmetry, as expected.

\subsection{Supersymmetry of the supergravity background}

In this section, we review the construction of the Killing spinors
of the background $AdS_4\times X_7$ when $X_7$ is a weak $G_2$ manifold.

Let us begin with the 11 dimensional supergravity field equations.
If we set gravitino to be $0$, the field equation becomes
\begin{align}
& R_{MN}=\frac{1}{2\times 3!}F_{MPQR}F_{N}{}^{PQR}
       -\frac{1}{6\times 4!}g_{MN}F_{PQRS}F^{PQRS},\nn\\
& \del_{M}\sqrt{-g}F^{MNPQ}+\frac{1}{2\times (4!)^2}\epsilon^{M_1\dots 
 M_8NPQ}F_{M_1\dots M_4}F_{M_5\dots M_8}=0,\nn\\
& (M,N,P,Q,\dots=0,1,\dots,10), \label{M-field-eq}
\end{align}
where $F$ is the 4-form field strength. The metric of $AdS_4\times X_7$
can be written as
\begin{align}
 ds^{2}=dr^{2}+e^{-4r}dx^{\mu}dx^{\nu}\eta_{\mu\nu}
  +g_{mn}dy^{m}dy^{n}, \quad (\mu,\nu=0,1,2,3,\quad m,n=4,\dots,10),
 \label{M-metric}
\end{align}
where $g_{mn}$ is the weak $G_2$ metric of $X_7$ which is normalised as
 $R_{mn}=6g_{mn}$. We set the vielbein $e^{a'},\ a'=0,1,2,3$
of $AdS_4$ space-time as
\begin{align}
  e^{3}=dr ,\  e^{p}=e^{-2r}dx^{p},\qquad (p=0,1,2),
\end{align}
and we denote the vielbein of $X_7$ as $e^{a},\ (a=4,\dots,10)$.
The solution of the 4-form can be written with these notations as
\begin{align}
 F=6e^{0}e^{1}e^{2}e^{3}.\label{M-4-form}
\end{align}
The solution (\ref{M-metric}),(\ref{M-4-form}) can have a parameter (``radius''),
but we set it $1$ as in the IIB case.\footnote{
Actually, we set the radius of $AdS_4$ to $\frac12$ in the metric
(\ref{M-metric}) for convenience. While, the ``radius'' of $X_7$ is $1$, which
means the normalisation $R_{mn}=6g_{mn}$.}

Now, let us turn to the Killing spinor analysis. The gravitino condition of
the SUSY parameter (Majorana spinor) $\epsilon$ in a bosonic background can be 
expressed as
\begin{align}
 \del_{A}\epsilon + \frac14 \omega_{A}{}^{BC}\Gamma_{BC}\epsilon
 + \frac{1}{288}F_{BCDE}\Gamma_{A}{}^{BCDE}\epsilon
  -\frac{1}{36}F_{ABCD}\Gamma^{BCD}\epsilon=0,\label{M-Killing1}
\end{align}
where $A,B,C,\dots=0,1,\dots,10$ is the 11 dimensional local Lorents indices,
and $\omega_{A}{}^{BC}$ is the torsion-free spin connection.

It is convenient to express the 11 dimensional gamma matrices $\Gamma^{A}$
as the form
\begin{align}
  \Gamma_{a'}=\gamma_{a'} \ox \id,\qquad (a'=0,1,2,3),\qquad
  \Gamma_{a}=\gamma_{(5)} \ox \gamma_{a},\qquad (a=5,\dots,11),
\end{align}
where $\gamma_{a'}\ a'=0,1,2,3$ are the $4 \times 4$ gamma matrices of SO(3,1)
and $\gamma_{a}\ a=4,\dots,11$ are the $8 \times 8$ gamma matrices of SO(7).
Here we also use the SO(3,1) chirality operator $\gamma_{(5)}:=i\gamma^{0123}$.
If we write the Killing spinor as $\epsilon=\lambda \ox \chi$,
the Killing spinor equation (\ref{M-Killing1}) reduces to
\begin{align}
  &\del_{a'}\lambda+\frac14 \omega_{a'}{}^{b'c'}\gamma_{b'c'}\lambda
  + i\gamma_{a'}\gamma_{(5)}\lambda = 0,\quad (a',b',c'=0,1,2,3),
  \label{AdS4-Killing}\\
 &\del_{a}\chi+\frac14 \omega_{a}{}^{bc}\gamma_{bc}\chi -\frac{i}{2}\gamma_{a}
  \chi =0,\quad (a,b,c=4,\dots,10).\label{real-Killing}
\end{align}
Eq.(\ref{real-Killing}) implies that $\chi$ is a real Killing spinor of
$X_7$. $X_7$ admits a real Killing spinor
if and only if $X_7$ is a weak $G_2$ manifold.

We need the $AdS_4$ part of the spin connection 
$\omega^{a'b'}=\omega_{c'}{}^{a'b'} e^{c'}$ in order to solve 
eq.(\ref{AdS4-Killing}). 
The non-zero components of this spin connection are 
$\omega^{3p}=-\omega^{p3}=2e^{p},\ p=0,1,2$. As in the IIB case, it is
convenient to decompose $\lambda=\lambda_+ + \lambda_-$ with
$i\gamma_3\gamma_{(5)}\lambda_{\pm}=\pm \lambda_{\pm}$. By using these notations,
the $AdS_4$ part of the Killing spinor equation (\ref{AdS4-Killing}) becomes
\begin{align}
 \frac{\del}{\del r} \lambda_{\pm}=\mp\lambda_{\pm},\qquad
 \frac{\del}{\del x^{p}} \lambda_{-}=0,\qquad
 \frac{\del}{\del x^{p}} \lambda_{+}+ 2 e^{-2r}\gamma_{3p}\lambda_{-} =0,\qquad
 p=0,1,2.
\end{align}
These equations are easily solved as
\begin{align}
 \lambda_{+}=e^{-r}(2x^{p}\gamma_{p3}\zeta_{-}+\zeta_{+}),\qquad
 \lambda_{-}=e^{r}\zeta_{-},\nn\\ \zeta_{\pm}:(\text{constant 4-component
 spinor}),\qquad i\gamma_{3}\gamma_{(5)}\zeta_{\pm}=\pm\zeta_{\pm}.
\end{align}

In summary, we obtained the Killing spinors in $AdS_4\times X_7$
\begin{align}
 &\epsilon=\lambda \ox \chi,\qquad
 \lambda=e^{-r}(2x^{p}\gamma_{p3}\zeta_{-}+\zeta_{+})
          + e^{r}\zeta_{-},
\end{align}
where $\zeta_{\pm}$ is constant 4-component spinors with  
$i\gamma_{3}\gamma_{(5)}$ eigenvalues $(\pm)$,
and $\chi$ is a real Killing spinor in the weak $G_2$ manifold $X_7$.
We consider M5-branes in this background in the next subsection.

\subsection{Supersymmetry of the brane background}
Let us introduce $AdS_3\times L_3$ M5-brane to the background considered above
for an associative submanifold $L_3$ of the weak $G_2$ manifold $X_7$.
For the supersymmetry in the presence of M5-brane, the associated Killing 
spinors $\epsilon$ must satisfy
\begin{align}
 \Gamma \epsilon = \epsilon,
\end{align}
where $\Gamma$ is the matrix for the kappa symmetry projection.
This gamma is expressed as
\cite{Howe:1997fb,Aganagic:1997zq,Pasti:1997gx,%
Bandos:1997ui,Cederwall:1998gg,Bandos:1997gm}
\begin{align}
 \Gamma:=\frac{1}{6!\sqrt{-\det \Gi}}\varepsilon^{j_1\dots j_6}\left[
         \Ga{j_1\dots j_6}+40\Ga{j_1j_2j_3} 
h_{j_4 j_5 j_6}\right].
\end{align}
We denote here by $h_{j_1j_2j_3}$ the 3-form field strength on the world-volume
which is self-dual with respect to the induced metric $\Gi$. We also use
the notation $\Ga{j_1\dots j_p}$ defined in eq.(\ref{Gamma2}).

The brane configuration considered here is as follows. We set the
world-volume coordinates as $\xi^{i},\ i=0,1,3,4,5,6$, and the configuration
is
\begin{align}
& x^{\alpha}=\xi^{\alpha},\quad (\alpha=0,1),\qquad r=\xi^{3},\qquad
x^{2}=\frac12 Me^{2r},\quad (M:\text{ constant}),\\
& y^{m}=y^{m}(\xi^4,\xi^5,\xi^6),\quad (\text{associative immersion}).
\end{align}
In this case, the induced metric becomes
\begin{align}
& ds^{2}_{\text{ind}}=(1+M^2)dr^2+e^{-4r}(-(dx^0)^2+(dx^1)^2)+
        G_{jk}d\xi^{j}d\xi^{k},\\
& G_{jk}=g_{mn}\deldel{y^{m}}{\xi^j}\deldel{y^{n}}{\xi^k},\qquad
 (j,k=4,5,6).
\end{align}
We also introduce the world-volume self dual 3-form field strength
\begin{align}
& h= \frac12 f (1+*_{\text{ind}})
     \sqrt{\det G}d\xi^{4}d\xi^{5}d\xi^{6},
\end{align}
where $f$ is a constant and $*_{\text{ind}}$ is
 the Hodge dual with respect to the induced metric.

In this brane background, the matrix $\Gamma$ becomes
\begin{align}
 \Gamma=-\frac{1}{\sqrt{1+M^2}}\Gamma_{01}(M\Gamma_{2}+\Gamma_{3})\frac{1}{\sqrt
 {\det G}}\Ga{456}-f\left[
  \frac{1}{\sqrt{1+M^2}}\Gamma_{01}(M\Gamma_{2}+\Gamma_{3})
 -\frac{1}{\sqrt{\det G}}\Ga{456}\right]. \label{M-Gamma1}
\end{align} 
The key formula for the analysis of this equation is
\begin{align}
 \deldel{y^{m_1}}{\xi^{4}}\deldel{y^{m_2}}{\xi^{5}}\deldel{y^{m_3}}{\xi^{6}}
e^{a_1}_{m_1}e^{a_2}_{m_2}e^{a_3}_{m_3}\gamma_{a_1a_2a_3}\chi
=-i\sqrt{\det G} \chi,
\label{M-key-formula}
\end{align}
for an associative immersion $(\xi^{4},\xi^{5},\xi^{6})\to X_7$
and a real Killing spinor $\chi$ of $X_7$.  The formula
(\ref{M-key-formula}) is proved in appendix \ref{app-M-key-formula}.
If we use eq.(\ref{M-key-formula}), the relation
$ \Ga{456}\epsilon=-i\sqrt{\det G} \gamma_{(5)}\epsilon $
is satisfied. Then, $\Gamma \epsilon$ reads
\begin{align}
 \Gamma \epsilon = \frac{1}{\sqrt{1+M^2}}i\Gamma_{01}(M\Gamma_{2}+\Gamma_{3})
\gamma_{(5)}\epsilon - f\left[
\frac{1}{\sqrt{1+M^2}}\Gamma_{01}(M\Gamma_{2}+\Gamma_{3})+
i\gamma_{(5)}\right]\epsilon.
\end{align}
At this stage, the problem is reduced to the $AdS_4$ part.
The relation eq.(\ref{M-Gamma1}) is equivalent to
\begin{align}
& \Gamma'\lambda=\lambda,\\
& \Gamma' =\frac{1}{\sqrt{1+M^2}}(M\gamma_3 - \gamma_2)
   + f \left[\frac{1}{\sqrt{1+M^2}}(M+\gamma_{32})-\gamma_{3}
  \right]i\gamma_{3}\gamma_{(5)},\\
&\lambda=e^{-r}(2x^{\alpha}\gamma_{\alpha 3}\zeta_{-}+\zeta_{+})
   + e^{r}(1+M\gamma_{23})\zeta_{-},\qquad (\alpha=0,1).
\end{align}
The Killing spinors satisfying this equation exist only when
$f$ and $M$ satisfy the relation
\begin{align}
 f=\frac{M}{1+\sqrt{1+M^2}}.
\end{align}
In this case, the surviving Killing spinors are parametrised by 
4-component spinors $\zeta_{\pm}$ satisfying
\begin{align}
 \gamma_2 \zeta_{\pm}=-\zeta_{\pm}.
\end{align}
Consequently, the number of the remaining supercharges are the same as expected;
half of the number of the ones of 3-dimensional $\Ncal=1$
superconformal symmetry.

\section{Conclusion and Discussion}
\label{Conclusion}
In this paper, we analyse supersymmetric AdS branes in the context of
AdS/dCFT correspondence.
We construct the string backgrounds for superconformal defects.
Especially, we consider the $AdS_4 \times T^2$ brane background and
the corresponding defect CFT. We show that there are the RG flow from
$T^2$ defect to $S^2$ defect. We also construct the M-theory AdS/dCFT
backgrounds.

Describing the corresponding defect CFT to the $AdS_4\times T^{2}$ brane is an
interesting problem. This defect CFT will be obtained by
considering the low energy theory on the D3-brane at the singularity of
the special Lagrangian D5-brane cone. The singular nature of the D5-brane cone
is essential in this case. However, it also make this problem difficult.
One possible approach is to describe the $T^2$ defect as the low energy
theory of a known defect. A candidate for this high energy theory is
the defect field theory with a certain number of hyper multiplets.
This high energy theory corresponds to $AdS_4\times (S^2\cup S^2 \cup \dots)$.
Consider a set of special Lagrangian planes intersecting
at the origin, and deform it so that the tangent cone of the remaining 
singularity is the $T^2$ cone. If one can do this, one find
the high energy conformal defect and the operator of the appropriate
relevant deformation.

It will be interesting to consider the AdS brane in the decoupling limit from
the gravity theory \cite{Berman:2001fs,Suryanarayana:2003}.
This produces the AdS/CFT correspondence only in open strings
(or open membranes). Especially, in the case of $AdS_3 \times L_3$ M5-brane,
the resulting CFT becomes 2-dimensional one, and the conformal
symmetry will enhance to infinite dimensions like in
the case of ref. \cite{Brown:1986nw}.

Another interesting related problem is the string theory in AdS in the
small radius limit\cite{Tseytlin:2002gz,Karch:2002vn,Dhar:2003fi,Clark:2003wk}.
In this limit, the defect CFT becomes weak coupling and perturbation can be
used. In contrast,
the supergravity or DBI description becomes wrong in this limit, 
and the string correction is essential.
To consider AdS/dCFT correspondence in this limit will be
useful for understanding the strong stringy effect of the string theory.

\subsection*{Acknowledgement}
I would like to thank Tohru Eguchi, Yosuke Imamura and Yuji Sugawara for useful 
discussions and comments.

This work is supported in part by JSPS Research Fellowships
for Young Scientists.

\appendix

\section{Sasaki-Einstein manifolds and special Legendrian submanifolds}
\subsection{Definitions and properties}
Let us consider a $(2k-1)$-dimensional Sasaki-Einstein manifold $X$
with metric $g_{mn},\quad
(m,n=2,3,\dots,2k)$. We denote the coordinates of $X$ by $y^{m},\ (m=2,3,\dots,2k)$.
We fix the normalisation of the metric as
\begin{align}
 R_{mn}=(2k-2)g_{mn}.
\end{align}

$X$ has 3 special differential forms:
1-form $\theta$, $(k-1)$-form $A$ and $B$.
These forms satisfy
\begin{align}
& \nabla_{m} \theta_{n}+\nabla_{n} \theta_{m}=0,\qquad
 g^{mn}\theta_m\theta_n=1,\qquad
 \nabla_{p}(d\theta)_{mn}=-2g_{pm}\theta_{n}+2g_{pn}\theta_{m},\\
&\nabla_{m}(A+iB)_{npq}=\nabla_{[m}(A+iB)_{npq]},
\qquad d(A+iB)=i^k k*(A+iB).
\end{align}

The cone $C(X)$ over a Sasaki-Einstein manifold $X$ is a Calabi-Yau manifold.
We introduce the radial coordinate $r$,
and write the metric of the Calabi-Yau manifold as
\begin{align}
 ds_{CY}^2=dr^2+r^2 g_{mn}dy^{m}dy^{n}.
\end{align}
Then K\"ahler form $J_{MN}$ and holomorphic $k$-form
$\Omega_{MN\dots}$, $(M,N,\dots=1,2,\dots,2k)$ are expressed as
\begin{align}
 J = r dr \theta+\frac12 r^2 d\theta,\qquad
 \Omega=r^{k-1}dr (A+iB) + \frac{1}{k}r^{k} d (A+iB),
\end{align}
where $*_{2k-1}$ means $(2k-1)$ dimensional Hodge dual with respect to
the metric $g_{mn}$.

Sasaki-Einstein manifold $X$ has two linearly independent real Killing spinors.
A Real Killing spinor $\chi$ satisfies
\footnote{The difference of the appearance of the 
definition of ``real'' Killing spinor between this paper and \cite{Bar:1993}
is due to the convention of gamma matrices.}
\begin{align}
& \del_{c}\chi +\frac14 \omega_{c}{}^{ab}\gamma_{ab} \chi
 -\frac{i}{2}\alpha \gamma_{c} \chi=0,\qquad \alpha=\pm 1,\label{RK1}\\
& \del_{c}\epsilon:=(e^{-1})^{m}_{c}\deldel{}{y^m},\qquad 
 e^{a}_{m}:(\text{vielbein}),\qquad (e^{-1})^{m}_{c}e^{a}_{m}=\delta^{a}_{c},\\
&\omega_{c}{}^{ab}:\text{(spin connection)},\qquad
\gamma_{c}:\text{(gamma matrices)},\qquad \{\gamma_{a},\gamma_{b}\}=2\delta_{ab}.
\end{align}
If $k=$(odd), one real Killing spinor $\chi_{+}$ satisfies (\ref{RK1})
for $\alpha=+1$ and the other $\chi_{-}$ satisfies (\ref{RK1})
for $\alpha=-1$. If $k=$(even), both of the real Killing spinors satisfy 
(\ref{RK1}) for $\alpha=+1$. This $\alpha=\pm 1$ corresponds to the chirality
of the parallel spinor in the Calabi-Yau cone $C(X)$.

The $(k-1)$-dimensional submanifold $L$ in $X$ is
called special Legendrian submanifold if
\begin{align}
 A|_{L}=\vol L, \label{SLeg}
\end{align}
where $A|_{L}$ is the pullback of $A$ to $L$, and ${\rm vol} L$ is the volume 
form of $L$. The definition (\ref{SLeg}) is equivalent to
\begin{align}
 B|_{L}=\theta|_{L}=0.
\end{align}
There is another equivalent definition. Consider the cone $\tilde L$ of $L$.
 \begin{align}
  \tilde L=\{(r, y)\in C(X)|r \in \Rb_{+}, y\in L \subset X\}.
 \end{align}
$L$ is a special Legendrian submanifold in $X$, if and only if $\tilde L$
is a special Lagrangian submanifold in $C(X)$.

Some properties of special Legendrian submanifolds are also explained in recent 
papers \cite{Mikhailov:2002ya, Wang:2002}.

\subsection{Proof of the formula (\ref{IIB-key-formula})}
\label{app-IIB-key-formula}

We denote the vielbein of $C(X)$ by $E^{A},\ A=1,\dots,2k$ and gamma matrices
$\Gamma^{A}$.
Let us first prove the formula
\begin{align}
 \frac{1}{k!}E^{A_1}\dots E^{A_k}\Gamma_{A_1 \dots A_k} \psi_{\pm}^{c}|_{\tilde L}
   =\psi_{\pm} \vol \tilde L, 
\label{CYkey}
\end{align}
for a special Lagrangian submanifold $\tilde L$ in
a Calabi-Yau manifold $C(X)$, and parallel spinors $\psi_{\pm}$, when we adjust 
the phase of these spinors appropriately.
Since, the formula (\ref{CYkey}) is invariant under the
local rotation, we can show eq.(\ref{CYkey}) in the convenient local frame.
We set the local frame so that the holomorphic $k$-form and K\"ahler form
can be written as
\begin{align}
 \Omega=(E^{1}+iE^{k+1})(E^{2}+iE^{k+2})\dots(E^{k}+iE^{2k}),\qquad
 J=\sum_{j=1}^{k}E^{j}E^{k+j}.
 \label{Omega}
\end{align}
In this frame, the two linearly independent parallel spinors $\psi_{\pm}$
is constant spinors defined by
\begin{align}
&(\Gamma^{j}-i\Gamma^{j+k})\psi_{+}=0,\quad (j=1,\dots,k)
  \qquad \psi_{-}=(\Gamma^{1}+i\Gamma^{k+1})(\Gamma^{2}+i\Gamma^{k+2})\dots
    (\Gamma^{k}+i\Gamma^{2k})\psi_{+}.
\end{align}
By the explicit calculation, we can show
\begin{align}
& \frac{1}{k!}E^{A_1\dots A_k} \Gamma_{A_1\dots A_k}\psi_{+}=
  \psi_{-}\Omega+(\text{terms proportional to } J), \label{CYvol1}\\
& \frac{1}{k!}E^{A_1\dots A_k} \Gamma_{A_1\dots A_k}\psi_{-}=
  (-1)^{k(k-1)/2} \psi_{+}\bar{\Omega}+(\text{terms proportional to } J).
\label{CYvol2}
\end{align}
Since the charge conjugation satisfies $(\psi^{c})^{c}=(-1)^{k(k-1)/2}\psi$,
we can adjust the phase of $\chi_{+}$ so that
the charge conjugation becomes
\begin{align}
 \psi_{+}^{c}=(-1)^{k(k-1)/2}\psi_{-},\qquad \psi_{-}^{c}=\psi_{+}.
\label{CYcc}
\end{align}
If we use the equations (\ref{CYvol1}),(\ref{CYvol2}),(\ref{CYcc}),
and the facts $J|_{\tilde L}=\im \Omega|_{\tilde L}=0$,
 $\re \Omega |_{\tilde L}=\vol L$
for a special Lagrangian submanifold $\tilde L$,
we obtain the formula (\ref{CYkey}).

In order to prove the formula (\ref{IIB-key-formula}) from (\ref{CYkey}),
let us introduce the following form of the vielbein and gamma matrices.
\begin{align}
  & E^{1}=dr,\quad E^{a}=r e^{a},\quad (a=2,\dots,2k),\nn\\
 &\Gamma_{1}=1\ox \sigma_{1},\qquad
 \Gamma_{a}=\gamma_{a}\ox \sigma_{2},\qquad
 \Gamma_{(5)}=1 \ox \sigma_{3},\nn\\
 &\gamma_{a}:((2k-1)\text{dimensional gamma matrices}),\qquad \sigma_{1,2,3}:
(\text{Pauli matrices}). \label{cone-gamma}
\end{align}
We consider $k=3$ case (which is relevant to our case) below in this section.
The parallel spinors $\psi_{\pm}$ of $C(X)$ can be written in
terms of the real Killing spinors $\chi_{\pm}$ as
$\psi_{-}=\chi_{-}\ox \binom{0}{1},\ \psi_{+}=\chi_{+}\ox \binom{1}{0}$.
This definition and eq.(\ref{CYkey}) read
\begin{align}
 \frac12 e^{a}e^{b}\gamma_{ab} \chi_{-}^{c}|_{L}
 =\chi_{-} \vol L.
\end{align}
If we write this equation by the world-volume coordinate $\xi^{5,6}$,
we obtain the formula (\ref{IIB-key-formula}) (We write in this appendix
$L$ and $X$ instead of $L_2$ and $X_5$).

\section{Weak $G_2$ manifolds and associative submanifolds}

\subsection{Definitions and properties}
Let us consider a 7-dimensional weak $G_2$ manifold $X_7$
with metric $g_{mn},\quad
(m,n=2,3,\dots,8)$. We denote the coordinates of $X_7$
by $y^{m},\ (m=2,3,\dots,8)$.
We fix the normalisation of the metric as
\begin{align}
 R_{mn}=6 g_{mn}.
\end{align}

Weak $G_2$ manifold $X_7$ has a special 3-form $\Phi$, which satisfies
\begin{align}
 d\Phi=4 *\Phi.
\end{align}

The cone of $X_7$ becomes a Spin(7) holonomy manifold $C(X_7)$. The metric is
\begin{align}
 ds_{\rm Spin (7)}^{2}=dr^2+r^2g_{mn}dy^m dy^n.
\end{align}
The Cayley 4-form $\Psi$ of Spin(7) holonomy manifold $C(X_7)$ becomes
\begin{align}
 \Psi=r^3dr \Phi+\frac14 r^4 d\Phi
\end{align}

The weak $G_2$ manifold $X_7$ has a real Killing spinor $\chi$ which satisfies
\begin{align}
& \del_{c}\chi +\frac14 \omega_{c}{}^{ab}\gamma_{ab} \chi
 -\frac{i}{2}\gamma_{c} \chi=0, \qquad\label{RKG}
\end{align}
where we use the same notation as eq.(\ref{RK1}).

A 3-dimensional submanifold $L_3$ of $X_7$ called associative submanifold
if
\begin{align}
 \Phi|_{L_3}=\vol L_3.
\end{align}
An associative submanifold in $X_7$
is related to a Cayley submanifold in $C(X_7)$.
Consider the cone $\tilde L_4$ of $L_3$ in $C(X_7)$. Then,
$L_3$ is an associative submanifold in $L_3$ if and only if $\tilde L_4$ in
 $C(X_7)$ is an Cayley submanifold.

\subsection{Proof of the formula (\ref{M-key-formula})}
\label{app-M-key-formula}
We denote the vielbein of $C(X_7)$ by $E^{A},\ A=1,\dots,8$ and gamma matrices
$\Gamma^{A}$.
Let us first prove the formula
\begin{align}
 \frac{1}{4!}E^{ABCD}\Gamma_{ABCD} \psi|_{\tilde L_4}
   =\psi \vol \tilde L_{4}, 
\label{7key}
\end{align}
for a Cayley submanifold $\tilde L_4$ in
a Spin(7) holonomy manifold $C(X_7)$, and a parallel spinor $\psi$.
We use in eq.(\ref{7key}) the notation $E^{AB\cdots}=E^{A}E^{B}\cdots$.
Since, the formula (\ref{7key}) is invariant under
local rotation, we can show eq.(\ref{7key}) in the convenient local frame.
We take the local frame in which the Cayley 4-form $\psi$ can be written in
the standard form
\begin{align}
 \Psi=&E^{1234}+E^{1256}+E^{1278}+E^{1357}-E^{1368}
    -E^{1458}-E^{1467} \nn\\& -E^{2358}-E^{2367}-E^{2457}
    +E^{2468}+E^{3456}+E^{3478}+E^{5678},\label{74}
\end{align}
where $E^{ABCD}:=E^{A}E^{B}E^{C}E^{D}$. In this frame, the parallel
spinor $\psi$ is the constant spinor satisfying
\begin{align}
 \psi=\Gamma^{1234}\psi=\Gamma^{1256}\psi
     =\Gamma^{1357}\psi=\Gamma^{12345678}\psi.\label{7S}
\end{align}
By local Spin(7) transformation, we can take the frame in which
the tangent space of Cayley submanifold $\tilde L_4$ is spanned by
$E^1,E^2,E^3,E^4$. Note that local Spin(7) transformation does not change
the form (\ref{74}) and (\ref{7S}). In this frame, left hand side of (\ref{7key})
becomes
\begin{align}
 \frac{1}{4!}E^{ABCD}\Gamma_{ABCD} \psi|_{\tilde L_4}&
  =E^{1234}\Gamma_{1234} \psi
=\psi \vol \tilde L_4.
\end{align}
This is the formula (\ref{7key}).

In order to show eq.(\ref{M-key-formula}) from (\ref{7key}), we take
the frame and gamma matrices of (\ref{cone-gamma}). 
The parallel spinor $\psi$ can be written with the real Killing spinor $\chi$
 as $\psi=\chi\otimes \binom{1}{0}$ in these notations.
Then, eq.(\ref{7key}) reduces to
\begin{align}
 \frac{1}{3!}e^{a}e^{b}e^{c}\gamma_{abc}\chi|_{L_3}
  =-i\chi \vol L_3.
\end{align}
If we rewrite this equation by world-volume coordinate,
we obtain the formula (\ref{M-key-formula}).

\providecommand{\href}[2]{#2}\begingroup\raggedright\endgroup

\end{document}